\documentstyle[preprint,tighten,eqsecnum,aps,floats,psfig,epsfig]{revtex}

\newcommand{\fd}{fluc\-tu\-a\-tion-dis\-si\-pa\-tion }
\newcommand{\beq}{\begin{equation}}
\newcommand{\beqa}{\begin{eqnarray}}
\newcommand{\eeq}{\end{equation}}
\newcommand{\eeqa}{\end{eqnarray}}
\newcommand{\eq}{{\rm eq}}
\newcommand{\dpar}{\partial}
\renewcommand{\d}{{\rm d}}

\begin{document}
\draft

\title{
Aging in ferromagnetic systems at criticality near four dimensions
}
\author{Pasquale Calabrese and Andrea Gambassi}
\address{Scuola Normale Superiore and INFN,
Piazza dei Cavalieri 7, I-56126 Pisa, Italy. 
\\
{\bf e-mail: \rm
{\tt calabres@df.unipi.it},
{\tt andrea.gambassi@sns.it}
}
}

\date{\today}

\maketitle

\begin{abstract}
We study the off-equilibrium response and correlation functions and the 
corresponding \fd ratio for a purely dissipative relaxation of an $O(N)$
symmetric vector model~(Model A) below its upper critical dimension.
The scaling behavior of these quantities is analyzed and the associated 
universal functions
are determined at first order in $\epsilon=4-d$ in the high-temperature phase 
and at criticality. A non trivial
limit of the \fd ratio is found in the aging regime
$\displaystyle X^\infty = {1\over 2}\left(1-{\epsilon\over 4} {N+2 \over N+8}\right)+O(\epsilon^2)$.

\end{abstract}



\section{Introduction}
\label{intr}

The time evolution of a system relaxing towards equilibrium is characterized
by two different regimes: a transient behavior with off-equilibrium evolution, 
for $t<t_\eq$, and a stationary equilibrium evolution for $t>t_\eq$.
In the former a dependence of the behavior of the system on initial conditions
is expected, while in the latter homogeneity of time and time reversal symmetry
(at least in absence of external fields) are recovered; dynamics of 
fluctuations are thus described in terms of ``equilibrium'' dynamics, with
a characteristic time scale diverging at the critical 
point~(critical slowing down)~\cite{HH}.

Consider a ferromagnetic model in a disordered state for the initial time 
$t=0$, and quench it at its critical temperature.
During the relaxation a small external field $h$ is applied at ${\bf x}=0$ 
after a waiting time $s$. At time $t$ the order parameter response to $h$ is 
given by the response
function $R_{\bf x} (t,s)=\delta\langle \phi_{\bf x} (t)\rangle/ \delta h (s)$,
where $\phi$ is the order parameter and $\langle \cdot \rangle$
stands for the mean over stochastic dynamics.
Since the system does not reach the equilibrium this function will 
depend both on $s$~(the ``age'' of the system) and $t$.
This behavior is usually referred to as aging and was
firstly noted in spin glass systems~\cite{review}.

To characterize the distance from equilibrium of an aging system, 
evolving at a fixed temperature $T$, the 
\fd ratio~(FDR) is usually introduced \cite{ckp-94,ck-94}:
\beq
X_{\bf x}(t,s)=\frac{T\, R_{\bf x}(t,s)}{\dpar_s C_{\bf x}(t,s)} \; ,
\label{dx}
\eeq
where $C_{\bf x}(t,s)=\langle \phi_{\bf x}(t)\phi_{\bf 0}(s)\rangle$, the 
two-time correlation function.
When the waiting time $s$ is greater than $t_\eq$ the dynamics is
homogeneous in time and the \fd theorem leads to $X_{\bf x}(t,s)=1$.
This is no longer true in the aging regime \cite{ckp-94}.

In recent years, several works
\cite{review,ckp-94,ck-94,nb-90,cd-95,ckp-97,barrat,bbk-99,fmpp-98,zkh-00,cst-01,clz}
have been devoted to the study of the FDR for systems exhibiting domain 
growth~\cite{bray},
or for aging systems such as glasses and spin glasses,
showing that in the low-temperature phase
$X(t,s)$ turns out to be a non-trivial function of its two arguments.
In particular analytical and numerical studies indicate that the limit 
\beq
X_{{\bf x}=0}^\infty=\lim_{s\to\infty}\lim_{t\to\infty}X_{{\bf x}=0}(t,s),
\label{xinfdef}
\eeq
vanishes throughout the low-temperature phase both for spin glass and simple
ferromagnetic system~\cite{ckp-97,barrat,bbk-99,zkh-00,cst-01}.

Only recently \cite{ckp-94,lz-00,gl-000,gl-00,bhs-01,gl-02b} 
attention has been paid 
to the FDR, for non-equilibrium, non-disordered, and unfrustrated systems  at 
criticality.
It has been argued that the FDR~(\ref{xinfdef}) is a novel universal quantity 
of non-equilibrium critical dynamics. 
Correlation and response functions were exactly computed in the simple cases
of a random walk, a free Gaussian field, and a two-dimensional XY model at 
zero temperature and the value $X_{{\bf x}=0}^\infty=1/2$ was found \cite{ckp-94}.
The same problem has been addressed for the 
$d$-dimensional spherical model \cite{gl-00}, 
for the one dimensional Ising-Glauber chain \cite{lz-00,gl-000}
and Monte Carlo simulation was done for the 
two dimensional Ising model \cite{gl-00}. 
In all cases $X_{{\bf x}=0}^\infty$ has values ranging 
between $0$ and $1\over 2$ while for some urn models  
 a different range has been found~\cite{gl-02}.
Also the scaling form for $R_{\bf x=0}(t,s)$ was rigorously established 
using conformal invariance \cite{hpgl-01}.

In this work we investigate the non-equilibrium correlation
and response functions and the associated FDR for the $O(N)$
ferromagnetic model with purely dissipative relaxation 
dynamics~(Model A  of Ref.~\cite{HH}) both at the critical point and in the 
high-temperature phase, using a field-theoretical approach~(never applied so 
far ---~see concluding remarks of Ref.~\cite{gl-00}),
at first order in an $\epsilon$-expansion.

The paper is organized as follows.
In Section \ref{sec2} we briefly introduce the model, the scaling forms and 
the Gaussian~(mean-field) result.
In Section \ref{sec3} we derive the $\epsilon$-expansion for the FDR
for all values of $s$ and $t$. 
Finally in Section \ref{disc} we summarize our results and discuss some
points needed further investigation.
In Appendix we report some useful details on the one loop calculation.

\section{The model}
\label{sec2}

Let us consider the purely dissipative relaxation dynamics of a $N$-component 
field $\varphi({\bf x},t)$
described by the stochastic Langevin equation~(Model A of Ref.~\cite{HH})
\beq
\label{lang}
\dpar_t \varphi ({\bf x},t)=-\Omega 
\frac{\delta \cal{H}[\varphi]}{\delta \varphi({\bf x},t)}+\xi({\bf x},t) \; ,
\eeq
where $\cal{H}[\varphi]$ is the Landau-Ginzburg Hamiltonian
\beq
{\cal H}[\varphi] = \int \d^d x \left[
\frac{1}{2} (\partial \varphi )^2 + \frac{1}{2} r_0 \varphi^2
+\frac{1}{4!} g_0 \varphi^4 \right] ,\label{lgw}
\eeq
$\Omega$ the kinetic coefficient, and 
$\xi({\bf x},t)$ a zero-mean stochastic Gaussian noise with
\beq
\langle \xi_i({\bf x},t) \xi_j({\bf x}',t')\rangle= 2 \Omega \, \delta({\bf x}-{\bf x}') \delta (t-t')\delta_{ij}.
\eeq  

The equilibrium correlation functions, generated by the Langevin 
equation~(\ref{lang}) and averaged over the noise $\xi$, 
can be obtained by means of the field-theoretical action \cite{bjw-76,ZJ-book} 
\beq
S[\varphi,\tilde{\varphi}]= \int \d t \int \d^dx 
\left[\tilde{\varphi} \frac{\partial\varphi}{\partial t}+
\Omega \tilde{\varphi} \frac{\delta \mathcal{H}[\varphi]}{\delta \varphi}-
\tilde{\varphi} \Omega \tilde{\varphi}\right].\label{mrsh}
\eeq
where $\tilde{\varphi}({\bf x},t)$ is the response field.

In Ref.~\cite{jss-89} this formalism was extended in order to incorporate
a macroscopic initial condition into Eq. (\ref{mrsh}):
one has also to average over the initial configuration 
$\varphi_0({\bf x})=\varphi(t=0,{\bf x})$ 
with a weight $e^{-H_0[\varphi_0]}$ given by
\beq
H_0[\varphi_0]=\int\! \d^d x\, \frac{\tau_0}{2}(\varphi_0({\bf x})-a({\bf x}))^2.
\eeq
This specifies an initial state $a({\bf x})$ with correlations proportional to
$\tau_0^{-1}$. In this way all response and correlation functions may
be obtained, following standard methods \cite{bjw-76,ZJ-book}, by a 
perturbative expansion of the functional weight
$e^{-(S[\varphi,\tilde{\varphi}]+H_0[\varphi_0])}$.

The propagators~(Gaussian two point correlation and response functions) 
of the resulting theory are \cite{jss-89} 
\beqa
\langle \tilde{\varphi_i}({\bf q},s) \varphi_j(-{\bf q},t) \rangle_0 =& 
\delta_{ij} R^0_q(t,s)=&\delta_{ij} \,\theta(t-s) G(t-s),\label{Rgaux}\\
\langle \varphi_i({\bf q},s) \varphi_j(-{\bf q},t) \rangle_0 =&
\delta_{ij} C^0_q(t,s)=& {\delta_{ij} \over q^2+r_0}\left[ G(|t-s|)+\left(\frac{r_0 +q^2}{\tau_0}-1
\right) G(t+s)\right], \label{Cgaux}
\eeqa
where
\beq
G(t)=\displaystyle{e^{-\Omega (q^2+r_0) t}} \label{GG}.
\eeq
It has also been shown that $\tau_0^{-1}$ is irrelevant~(in the 
renormalization group sense) for large times behavior \cite{jss-89}.  

\subsection{Scaling forms}

When a ferromagnetic system is quenched from a disordered initial state to
its critical point, the correlation length grows as $t^{1/z}$, where $z$ is
the dynamical critical exponents \cite{HH}. So in  momentum space, applying
standard scaling arguments, all the universal functions depend 
only on the two products $q^z \, t$ and $q^z\, s$.

In particular we expect the scaling forms \cite{jss-89}
\beqa
R_q(t,s)= q^{-2+\eta+z} \left( t\over s\right)^\theta F_R(\Omega q^z (t-s),t/s),
\label{Rsf}\\
C_q(t,s)= q^{-2+\eta} \left( t\over s\right)^\theta F_C(\Omega q^z (t-s),t/s),
\label{Csf}
\eeqa
where $\theta$ is the initial-slip exponent of response function related
to the initial-slip exponent of the magnetization $\theta '$ 
and to the autocorrelation exponent $\lambda_c$~\cite{huse-89}
by the relation \cite{jss-89}
\beq
\theta' =\theta +z^{-1}(2-z-\eta)=z^{-1}(d-\lambda_c).
\eeq

The functions $F_R(y,x)$ and $F_C(y,x)$ are universal apart the normalizations 
for small arguments. These functions are regular functions of both arguments, 
and for large $x$ they behave as:
\beqa
F_R(y,x)=F_R^\infty (y)+O(x^{-1}), \\
F_C(y,x)={F_C^\infty (y) \over x}+ O(x^{-2}),
\eeqa
so that, for $s \rightarrow 0$, these scaling forms reduce to ones of Ref.~\cite{jss-89}. 
We would also mention that transforming Eqs.~(\ref{Rsf},\ref{Csf}) 
in the real
${\bf x} $ space (to this end we have to assume that $F_R$ and $F_C$ 
are rapidly decreasing functions of $y$ for $y\rightarrow\infty$) and setting 
${\bf x}=0$ one  easily obtain the
scaling forms reported in Ref.~\cite{gl-00,hpgl-01}. They also
reduce to the equilibrium ones \cite{HH} when $t\sim s \gg 1$.

Let us introduce \cite{footnote}
\beq
{\cal X}_{\bf q}={\Omega R_{\bf q}(t,s)\over \dpar_s C_{\bf q}(t,s)}\label{Xq}
\eeq
then, from the scaling forms above is simple to show that, assuming 
$F_R(y,x)= O(y)$, $\forall x$, the FDR
may be written as
\beq
{\cal X}^\infty_{{\bf q}=0} = \lim_{s\to\infty}\lim_{t\to\infty}{\cal X}_{{\bf q}=0}(t,s)=
(1-\theta)^{-1} {F^\infty_R(0) \over F'^\infty_C (0)}.
\label{XFF}
\eeq

\subsection{Gaussian FDR}

For the Gaussian model we know exactly the response and correlations functions, so we can evaluate the FDR~(in \cite{ckp-94} the related quantity $X_{\bf x}$
has been considered, see Sec.~\ref{disc}).
From Eqs.~(\ref{Rgaux}, \ref{Cgaux}) and definition (\ref{Xq}) we have
\beq
\displaystyle
{\cal X}_q^0(t,s)=\left({\dpar_s C^0_q \over \Omega G^0_q}\right)^{-1}=
\left(1+e^{-2 \Omega (q^2+r_0) s}
+\Omega q^2 \tau_0^{-1} e^{-2 \Omega (q^2+r_0) s} \right)^{-1}.
\eeq

If the theory is off-critical ($r_0 \neq 0$) the limit of this ratio for 
$s\rightarrow \infty$ is $1$ for all values of $q$, in agreement with 
the idea that in the high-temperature phase all modes have a 
finite equilibration time, so that equilibrium is recovered and as a 
consequence the \fd theorem holds.
For the critical theory, i.e. $r_0 \propto T-T_c=0$, if $q\neq 0$
the limit ratio is again equal to one, whereas for $q=0$ we have
${\cal X}_{q=0}^0(t,s)=1/2$. This analysis clearly shows that the only mode 
characterized by aging, i.e. that ``does not relax'' to the equilibrium, is
the zero mode in the critical limit. 

\section{One-loop FDR}
\label{sec3}

The aim of this section is the computation of the non-equilibrium response and 
correlation functions for the purely dissipative dynamics of the $N$-vector 
model at one-loop order. 
We use here the method of renormalized field theory in the minimal 
subtraction scheme. 
The breaking of homogeneity in time gives rise to some technical problems in
the renormalization procedure in terms of one-particle irreducible correlation
functions~(see \cite{jss-89} and references therein) so our computation is done
in terms of connected functions.

\begin{figure}[t]
\centerline{\psfig{width=5truecm,file=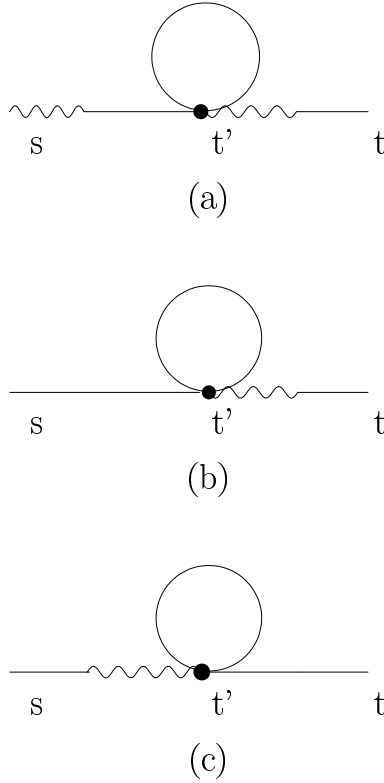}} 
\caption{Feynman diagrams contributing to the one-loop response~(a) and 
correlation function~((b)+(c)). 
Response functions are drawn as wavy-normal lines, whereas
correlators are normal lines. A wavy line is attached to the response field
and a normal one to the order parameter.}\label{fig}
\end{figure}

At one-loop order we have to evaluate, taking also into account causality 
\cite{bjw-76}, the three Feynman diagrams 
in Figure~\ref{fig}, one for the response function and two for
the correlation one. In terms of these diagrams we have
\beqa
R_q(t,s)= R_q^0(t,s)- {N+2 \over 6} g_0 (a) +O(g_0^2)\;,\nonumber\\
C_q(t,s)= C_q^0(t,s)- {N+2 \over 6} g_0 ( (b)+(c))+O(g_0^2)\;.\label{int}
\eeqa


In order to evaluate the FDR at criticality we have to set in
this perturbative expansion $r_0=0$~(massless theory).
We also set $\tau_0^{-1}=0$, since it is an irrelevant variable \cite{jss-89}, 
and $\Omega=1$ to lighten the notations.
The first step in the calculation of the
diagrams is the evaluations of the critical ``bubble'' $B_c(t)$, i.e. their 
common one-particle irreducible part.
We have, in generic dimension $d$,
\beq
B_c(t)= \int \! {\d^dq \over (2 \pi)^d} \; C_q(t,t)=-
{1\over d/2-1}{(2 t)^{1-d/2} \over (4 \pi)^{d/2}}=- N_d {\Gamma(d/2-1)\over 2^{d/2}} t^{1-d/2},
\eeq
where $\displaystyle N_d= {2\over (4 \pi)^{d/2} \Gamma(d/2)}$.
Note that the equilibrium contribution to $B_c(t)$ is zero for $d >2$.

Let us consider $t>s$ in the following. We may write
\beqa
(a)=\int_0^\infty \!\d t'\, R^0_q (t,t') B(t') R^0_q (t',s)\;, \nonumber\\
(b)=\int_0^\infty \!\d t'\, R^0_q (t,t') B(t') C^0_q (t',s)\;, \label{inte}\\
(c)=\int_0^\infty \!\d t'\, R^0_q (s,t') B(t') C^0_q (t',t)\;, \nonumber
\eeqa
where we set $r_0=0$ in $R^0_q$ and $C^0_q$.

Performing the integration and expanding in powers of $\epsilon$ we find
for the response function
\beq
R_q(t,s)=G(t-s) \left( 1+ \tilde{g}_0{N+2\over 24} \ln {t\over s} \right)+
O(\epsilon^2,\tilde{g}_0^2),\label{Reps}
\eeq
and for the correlation function
\beq
C_q(t,s)={G(t-s)-G(t+s)\over q^2} 
\left( 1+ \tilde{g}_0{N+2\over 24} \ln {t\over s} \right)-
\tilde{g}_0{N+2\over 24} {G(t+s) \over q^2} f(2 q^2\,s)+
O(\epsilon^2,\tilde{g}_0^2),\label{Ceps}
\eeq
where
\beq
f(v)=2\left[\int_0^v \! \d \xi \,\ln \xi \, e^\xi + (1-e^v)\ln v\right],
\label{f(x)}
\eeq
and $\tilde{g_0}=N_d g_0$. Note that $f(0)=0$, $f'(0)=-2$ and $f(v)$ has the following asymptotic expansion, for $v \gg 1$
\beq
f(v) = -2\, {e^v \over v}\left( 1 + {1 \over v} + {2 \over v^2} + \ldots + {k! \over v^k} + \ldots \right)\; .
\label{fas}
\eeq

In order to obtain the critical functions we have to set the renormalized 
coupling equal to its fixed point value.
At first order in $\epsilon$ \cite{ZJ-book}
\beq
\tilde{g}_0=\tilde{g}^*= {6\over N+8} \epsilon +O(\epsilon^2).
\eeq
Finally we get~(called $P_N={N+2 \over N+8}$)
\beq
R_q(t,s)=G(t-s) \left( 1+ \epsilon {P_N \over 4}  \ln {t\over s} \right)+
O(\epsilon^2)\;,
\eeq
\beq
C_q(t,s)={G(t-s)-G(t+s)\over q^2} 
\left( 1+ \epsilon {P_N \over 4} \ln {t\over s} \right)-
\epsilon {P_N \over 4} {G(t+s) \over q^2} f(2 q^2 s)+ O(\epsilon^2) \;,
\eeq
that are fully compatible with the scaling form given in the previous 
section, with
\beq
F_R(y,x) = e^{-y} + O(\epsilon^2) \; ,
\label{scalFR}
\eeq
and
\beq
F_C(y,x) = e^{-y} - \left[ 1 + \epsilon {P_N \over 4} 
f\left({2y \over x-1}\right)\right]e^{- y{ x+1 \over x-1}} + O(\epsilon^2) \; .
\label{scalFC}
\eeq
In particular we recognize the exponent 
$\theta=P_N \epsilon/4 + O(\epsilon^2)$ in 
agreement with Ref.~\cite{jss-89}, \mbox{$z = 2 + O(\epsilon^2)$}, 
$\eta = O(\epsilon^2)$ as expected, and that $F_R(y,x)$ is not affected by 
$O(\epsilon)$ corrections. It is also easy to find that 
\beq
F_C^\infty(y) = 2 y \left( 1 + \epsilon {P_N \over 2}\right) e^{-y} + O(\epsilon^2)\; .
\label{FCinf}
\eeq

Computing the derivative with respect to $s$ of the two-time correlation 
function and taking its ratio with the response function we have
\beq
{\cal X}^{-1}_q(s)=1+e^{-2q^2 s}-{P_N \epsilon \over 4} e^{-2q^2 s}
\left[{e^{2q^2 s}-1 \over q^2 s}-f(2 q^2 s)+ 2 f'(2 q^2 s)\right] 
+ O(\epsilon^2)\; .
\label{X-1} 
\eeq
Note that, at least at this order, the result is independent of the 
observation time $t$.
Using the large $v$ behavior of $f(v)$, cf. Eq.~(\ref{fas}), we find 
that the limit of the  FDR for $s \rightarrow \infty$ is equal to 1 for 
all $q \neq 0$.
Instead for $q=0$ we have~(using (\ref{f(x)}))
\beq
{\cal X}^\infty_{q=0}={1\over 2} \left(1 - \epsilon {P_N \over 4}\right) 
+ O(\epsilon^2)\; ,
\eeq
in agreement with Eq.~(\ref{XFF}) and with the scaling forms~(\ref{scalFR})
 and~(\ref{FCinf}).


Taking into account the effect of the mass $r_0$~(deviation from
critical temperature) in the previous computations, one obtains for the 
non-critical bubble~(contributing to the mass renormalization)
\beq
B(t)= N_d \left[{\pi \over 2 \sin d\pi/2} -
{1\over 2} \Gamma(d/2) \Gamma(1-d/2, 2r_0 t) \right] r_0^{d/2-1}\; ,
\label{massivebubble}
\eeq
where $\Gamma(x,y)$ is the incomplete $\Gamma$ function \cite{GR}.
Using this expression it is possible to determine, as previously done, correlation and response functions. We report the basic formulas in the Appendix. 
The final result is obtained computing the ratio ${\cal X}_q$ in terms of the renormalized parameters of the theory. 
It is then trivial, but algebraically cumbersome, to show that  ${\cal X}^\infty_q$ 
is equal to $1$ for all $q$ in the high temperature phase.


\section{Discussion}
\label{disc}

In this work we considered the off-equilibrium properties of the  purely 
dissipative relaxational dynamics of an $N$-vector model in the framework of 
field theoretical $\epsilon$-expansion.
We computed at first order in $\epsilon$ the FDR, as defined in (\ref{Xq})
as a function of the waiting time $s$ and of the observation
time $t$ both at criticality and in the high-temperature phase. 
The main result is that the ratio ${\cal X}^\infty_q$ is always $1$
unless at criticality for $q=0$, when it takes the value
\beq
{\cal X}^\infty_{{\bf q}=0}={1\over 2}\left(1-{\epsilon \over 4}{N+2 \over N+8}\right)
+ O(\epsilon^2)\; . 
\label{risf}
\eeq

To compare our result with some particular limit considered in the
literature \cite{ckp-94,gl-00} we have to relate this quantity to the analog 
in the real ${\bf x}$ space. The following heuristic argument 
may be useful to realize that the two ratios are exactly equal, i.e.
\beq
X^\infty_{{\bf x}=0}={\cal X}^\infty_{{\bf q}=0}\label{eq} \; .
\eeq
We may rewrite the FDR in real ${\bf x}$ space as a mean value of that
in momentum space with a weight given by $R_{\bf q}$:
\beq
X^{-1}_{{\bf x}=0} \equiv {\int \! \d^d q \, \dpar_s C_{\bf q}(t,s)
\over T \int \!\d^d q \, R_{\bf q}(t,s)}= 
{\int \! \d^d q\, R_{\bf q}(t,s) \, {\dpar_s C_{\bf q}(t,s) \over T R_{\bf q}(t,s)} 
\over \int \! \d^d q \, R_{\bf q}(t,s)}= 
{\langle {\cal X}^{-1}_{\bf q} \rangle}_{R_{\bf q}} \; .
\eeq
Now, since we expect $\displaystyle R_{\bf q} \propto e^{-q^2(t-s)}$, in the
limit $s,t\rightarrow \infty$~(in the right order) $X^{-1}_{{\bf x}=0}$
will take contributions only for the $q=0$ mode, i.e. apart a normalization,
the weight function $R_{\bf q}$ is a $\delta({\bf q})$. 
However we  note that at the first order in $\epsilon$ the equality
(\ref{eq}) is identically satisfied by our result since
$\displaystyle R_{\bf q} \propto e^{-q^2(t-s)}$~, cf. Eq.~(\ref{Reps}). 


In the limit $N\rightarrow \infty$ Eq.~(\ref{risf}) reduces to
$X^\infty=1/2-\epsilon/8 + O(\epsilon^2)$ that is the same as the expansion
 of the result for the spherical model near four dimension \cite{gl-00}.

A Monte Carlo simulation of the two dimensional Ising model gave for the
FDR the value $X^\infty=0.26(1)$ \cite{gl-00}, qualitatively in agreement
with our result for $\epsilon=2$, $X^\infty=5/12<1/2$.
To have a reliable quantitative prediction the knowledge of 
higher loop contributions is required.

Setting $\epsilon=1$ for $N=1$, one obtains $11/24$.
This number, that is a rough estimate of the actual three-dimensional value, 
may be measured in numerical or experimental works.

This work may be easily extended to more
realistic models than those previously considered in literature, 
contributing to the understanding of 
out-of-equilibrium dynamic phenomena,
currently under intensive investigation, by means of the
powerful tools of perturbative field theory. 


\section*{Acknowledgments}
We would thank S.~Caracciolo, A.~Pelissetto and E.~Vicari for a critical reading of the manuscript and useful comments. We acknowledge in a special way A.~Pelissetto for having introduced us to these problems.   

\appendix

\section*{Details of computations}

We summarize here the main analytical results useful for the computation of 
correlation and response functions for $r_0 \ge 0$\ at one loop.
Again one has to perform all the needed integrations over the times, as in 
Eq.~(\ref{inte}) with the free field correlator and response function~ 
(\ref{Rgaux},~\ref{Cgaux}). At variance with the critical theory a 
renormalization of the parameter $r_0$\ is now required 
 to cancel dimensional poles both in $R_q$ and $C_q$. 

Let us introduce the function
\beq
Y(t) \equiv \int_0^t \! \d\tau\, B(\tau)\; ,
\label{defY}
\eeq
and
\beq
W(t) \equiv \int_0^t \! \d\tau\, G(-2\tau)\, B(\tau)\; ,
\label{defW}
\eeq
where $G(t)$\ and $B(t)$ are given in Eq.~(\ref{GG}) 
and~(\ref{massivebubble}), respectively.
In terms of $Y$ and $W$ we obtain (for $t>s$\ and $\tau_0^{-1}=0$, $\Omega =1$)
\beq
(a)= G(t-s)[Y(t)-Y(s)]\; ,
\label{expa}
\eeq
and
\beqa
(b) + (c) = 
 {1 \over q^2 + r_0} \left\{ G(t-s)[Y(t)-Y(s)] - G(t+s)[Y(t)+Y(s)] + 2 G(t+s) W(s)\right\} \; .\nonumber \\
\label{expbplusc}
\eeqa
In the following with $Y$\ and $W$\ we mean also their analytic continuation 
in $d$.

An explicit computation leads to
\beq
Y(t) = \frac{r_0^{d/2-2} }{2 (4\pi)^{d/2}} \left\{ (2 r_0 t + d/2 -1)[\Gamma(1-d/2) - \Gamma(1-d/2, 2 r_0 t)] + (2 r_0 t)^{1-d/2} e^{-2 r_0 t}\right\} \; ,
\label{expY}
\eeq
and
\beqa
\lefteqn{W(t) = } \nonumber \\
&& ={ 1\over 2 (4\pi)^{d/2}} {r_0^{d/2-1}\over q^2 + r_0} \left\{ G(-2t) [\Gamma(1-d/2) - \Gamma(1-d/2,2 r_0 t)] 
- (q^2/r_0)^{d/2-1} \Delta(1-d/2, 2q^2t)\right\} \; , \nonumber \\
\label{expW}
\eeqa
where we introduced
\beq
\Delta(v,w) \equiv \int_0^w \! \d\tau \,\tau^{v-1}\, e^\tau 
\eeq
(for $v\le 0$\ its analytic continuation has to be considered).

Expanding (\ref{expY},\ref{expW}) in $\epsilon = 4 - d$, we obtain
\beq
2 (4\pi)^{d/2} Y(t) = -{2\over \epsilon}(2 r_0 t + 1) -
 (2 r_0 t + 1)[\gamma(2 r_0 t) 
 - \ln r_0 ] + 1 +{e^{-2r_0t}\over 2 r_0 t} + O(\epsilon)\; , 
\label{expexpY}
\eeq
and
\beqa
\lefteqn{2 (4\pi)^{d/2} (q^2 + r_0) W(t) =} \nonumber\\
&& = -{2\over \epsilon}[r_0 G(-2t) + q^2] + q^2 [\ln q^2 -\delta(2 q^2 t)] 
- r_0 G(-2t) [\gamma(2 r_0 t) - \ln r_0 ]  + O(\epsilon) \; , 
\label{expexpW}
\eeqa
where
\beq
\gamma(v) \equiv 1 + e^{-v} \left(\ln v + {1\over v}\right) +
 \int_0^v\! \d \xi\, \ln \xi\, e^{-\xi} \; , 
\eeq
and
\beq
\delta(v) \equiv 1 + e^v \left(\ln v - {1\over v}\right) 
- \int_0^v\! \d \xi\, \ln \xi\, e^\xi \; .
\eeq
It is easy to find that $f(v)$\ in Eq.~(\ref{f(x)}) is related to 
$\delta(v)$\ by 
\beq
f(v) = 2\left[1 + \ln v - \delta(v) - {e^v \over v} \right]\; .
\eeq
Plugging Eqs.~(\ref{expexpY}) and (\ref{expexpW}) into Eq.~(\ref{expa}) and (\ref{expbplusc}) and then into Eqs.~(\ref{int}) it is easy to realize that to cancel dimensional poles both in $R_q(t,s)$ and $C_q(t,s)$\ a renormalization of the bare mass $r_0$ is sufficient (at least in the case $\tau_0^{-1}=0$ we are considering)
\beq
r_0 = Z_r r \ \ \ \mbox{with}\ \ \ Z_r = 1 + {N+2\over 3}{g_0\over (4\pi)^{d/2}} {1\over \epsilon} + O(g_0^2)\; ,
\eeq
in agreement with what one would expect from the corresponding static field-theory (see, for instance, Ref.~\cite{ZJ-book}). All the previously stated results easily follow from explicit expressions given above.

\end{document}